\documentclass[preprint,eqsecnum,nofootinbib]{revtex4}
\usepackage{graphicx,amssymb,amsbsy,bm,amsmath,psfrag,dsfont}
\usepackage{hyperref}

\begin{document}

\newcommand{\be}{\begin{equation}}
\newcommand{\ee}{\end{equation}}
\newcommand{\bea}{\begin{eqnarray}}
\newcommand{\eea}{\end{eqnarray}}
\newcommand{\no}{\noindent}

\newcommand{\la}{\lambda}
\newcommand{\si}{\sigma}
\newcommand{\vp}{\mathbf{p}}
\newcommand{\vk}{\vec{k}}
\newcommand{\vx}{\vec{x}}
\newcommand{\om}{\omega}
\newcommand{\Om}{\Omega}
\newcommand{\ga}{\gamma}
\newcommand{\Ga}{\Gamma}
\newcommand{\gaa}{\Gamma_a}
\newcommand{\al}{\alpha}
\newcommand{\ep}{\epsilon}
\newcommand{\app}{\approx}
\newcommand{\uvk}{\widehat{\bf{k}}}
\newcommand{\OM}{\overline{M}}

\title{Sterile Neutrinos and Light Dark Matter Save Each Other}
\author{Chiu Man Ho and Robert J. Scherrer}
\affiliation{Department of
  Physics and Astronomy, Vanderbilt University, Nashville, Tennessee
  37235, USA}
\date{\today}

\begin{abstract}
Short baseline neutrino experiments such as LSND and MiniBooNE seem to suggest the existence of light sterile neutrinos.
Meanwhile, current cosmic microwave background (CMB) and big bang
nucleosynthesis (BBN) measurements place an upper bound on the effective number of light
neutrinos, $N_{eff}$ and the PLANCK satellite will measure $N_{eff}$ to a much
higher accuracy and further constrain the number of sterile neutrinos allowed.
We demonstrate that if an MeV dark matter particle couples more strongly to electrons and/or photons than to neutrinos,
then $p$-wave annihilation after neutrino decoupling can reduce the value of $N_{eff}$
inferred from BBN and PLANCK. This mechanism can accommodate two eV sterile
neutrinos even if PLANCK observes $N_{eff}$ as low as the standard model theoretical value of 3.046, and a large neutrino
asymmetry is not needed to obtain the correct primordial element abundances.
The dark matter annihilation also weakens the cosmological upper bounds on the neutrino
masses, and we derive a relationship between the change in these bounds and the corresponding change in
$N_{eff}$. Dark matter with an electric dipole moment or anapole moment is a natural candidate that exhibits
the desired properties for this mechanism. Coincidentally, a dark matter particle with these properties and
lighter than 3 MeV is precisely one that can explain the 511 keV gamma-ray line
observed by INTEGRAL. We show that the addition of two eV sterile neutrinos
allows this kind of dark matter to be lighter than 3 MeV, which is otherwise ruled out by the CMB bound on
$N_{eff}$ if only active neutrinos are considered.

\end{abstract}
\maketitle

\section{Introduction}
\label{sec:intro}

The standard three-neutrino paradigm has been very successful in describing the oscillation phenomenon associated with
solar, atmospheric, reactor and accelerator neutrinos. However, in recent years, there have been some hints of deviations from this
three-neutrino paradigm in short baseline neutrino experiments such as LSND \cite{LSND} and MiniBooNE \cite{MiniNull,Mini}.

In LSND, the transition probability of $\bar{\nu}_{\mu} \rightarrow \bar{\nu}_{e}$ was measured, but the data indicated a 3.8$\sigma$ excess
of $\bar{\nu}_e$ events. Since only antineutrinos were involved in LSND, it was then noticed that if CPT is violated in the neutrino sector
such that the mass-squared splittings for neutrinos and antineutrinos are different, all the data could be reconciled \cite{LSNDCPTV}.
But this scheme was contradicted by the KamLAND \cite{KamLAND} data. Another CPT-violating scheme was then proposed by \cite{Lykken}, but
it was not compatible with a three-neutrino global analysis of the neutrino data \cite{Schwetz}.


Meanwhile, the idea of sterile neutrinos \cite{sterile} has also been invoked to resolve the LSND anomaly. But it turns out that adding only
one eV sterile neutrino to the standard three-neutrino picture, namely the (3+1) model, is not enough \cite{1sterile}. The (3+2) model
(with two eV sterile neutrinos) was proposed and shown to fit the data much better than the (3+1) model \cite{2sterile}.

The purpose of MiniBooNE was to confirm or exclude LSND. In the first release of data, they found no evidence for an excess of $\nu_e$ in
the $\nu_{\mu} \rightarrow \nu_{e}$ search \cite{MiniNull}. But their later search for $\bar{\nu}_\mu \rightarrow \bar{\nu}_e$ appeared to be consistent with LSND \cite{Mini}.
In contrast to their previous $\nu_{\mu} \rightarrow \nu_{e}$ search, MiniBooNE's latest result indicates that the neutrino and antineutrino data are consistent with each other \cite{LatestMiniBooNE}. However, the tension between appearance and disappearance (such as CDHS \cite{CDHS} and Bugey \cite{Bugey}) short baseline experiments remains \cite{Conrad}. A (3+3) model has been considered by \cite{Conrad2}, but the global fit to appearance and disappearance experiments is still poor.

Recently, a further hint of sterile neutrinos has emerged from a theoretical re-evaluation of the expected mean reactor antineutrino
flux emitted from nuclear reactors \cite{Mueller}. The new prediction suggests a flux 3\% higher than what was previously assumed
\cite{Schreckenback}. This implies that all reactor neutrino experiments with a baseline smaller than 100 m have measured a deficit
in the $\bar{\nu}_e$ events. Motivated by this new observation, the global short baseline neutrino oscillation data has been re-analyzed
in the (3+1) and (3+2) models \cite{Joachim}. It is shown that while the (3+1) model is still insufficient, the global fit improves
significantly in the (3+2) model. The conclusion is that the short baseline neutrino experiments now favor two sterile neutrinos with
the best-fit masses being $m_4=$ 0.68 eV and $m_5=$ 0.94 eV \cite{Joachim}.

While short baseline neutrino experiments suggest the existence of eV sterile neutrinos, there are useful constraints from cosmology.
For instance, the bound from Big Bang Nucleosynthesis (BBN) is quite stringent. The existence of extra relativistic degrees of freedom,
if thermalized, will increase the Hubble expansion rate and cause the weak interactions to freeze out at earlier time (higher temperature).
This will increase the neutron-to-proton ratio and may result in an overproduction of primordial helium and deuterium. The recent analyses
lead to $N_{eff} < 4.26$ \cite{Raffelt} and $N_{eff} < 4.1$ \cite{Serpico} at 95\% C.L. limit. The best fit derived by \cite{Raffelt} is
$N_{eff} =3.86$. As a result, while one fully thermalized eV sterile neutrino is preferred by BBN, two fully thermalized eV sterile neutrinos
are disfavored. This bound could be circumvented if there is a positive $\nu_e$ degeneracy parameter (or equivalently, neutrino asymmetry), $\xi_e$,
which reduces the neutron-to-proton ratio \cite{Foot,Danny}.
With a non-zero $\nu_e$ degeneracy parameter, the neutron-to-proton ratio is given by $n/p = \exp\{-(m_n-m_p)/T-\xi_e\}$. So the purpose of the $\nu_e$ degeneracy parameter is to reduce $n/p$ which has been elevated by the two thermalized eV sterile neutrinos.
In order to accommodate two eV sterile neutrinos, one would need the degeneracy parameter  $\xi_e$ to satisfy $ 0.03 < \xi_e < 0.14$ at 95\% C.L. with the best-fit value being $\xi_e =0.064$ \cite{Raffelt}.

In fact, if \,$\xi_e$\, is really as large as approximately 0.1, the active-sterile mixing will be delayed through the suppression of
the medium mixing angle \cite{Foot}. In this case, the medium mixing angle will be suppressed by the matter effect
which is dominated by the neutrino asymmetries \cite{Notzold}. The active-sterile transition probability is even suppressed by the
so-called quantum Zeno effect near the MSW resonance \cite{Me}. Thus, the resulting effect is that the active-sterile mixing will occur only
after neutrino decoupling and the sterile neutrinos will not be thermalized during the epoch of BBN. This understanding has recently been
confirmed by \cite{Hannestad} which solves the full quantum kinetic equations \cite{Stodolsky}. As a consequence, the effective number of neutrinos remains at $N_{eff} = 3.046$. Without the thermalized sterile neutrinos, the existence of $\xi_e$ would reduce $n/p$ and the question is how much
$\xi_e$ can be tolerated by BBN. It turns out that the answer depends on the value of the mixing angle $\theta_{13}$. Recently, DAYA-BAY \cite{DAYA-BAY} and RENO \cite{RENO} have observed $\sin^2\,(\,2 \theta_{13}\,) = 0.092 \pm 0.016\, (\textrm{stat.}) \pm 0.005\, (\textrm{syst.})$
and $\sin^2\,(\,2 \theta_{13}\,) = 0.113 \pm 0.013\, (\textrm{stat.}) \pm 0.019 \,(\textrm{syst.})$ at 68\% C.L.
respectively. Taking into account all of the measured mixing angles for the three active neutrinos and the effect of neutrino oscillations, it has been shown that $\xi_e \sim 0.1$ is indeed allowed by BBN \cite{Pastor}.

The CMB observations \cite{WMAP7B,Dunkley,Keisler} also place bounds on the effective number of neutrino degrees
of freedom $N_{eff}$. (See \cite{Bashinsky} for a discussion of the effect of $N_{eff}$ on the CMB fluctuations.)
At recombination, the temperature of photons $T_\gamma$ and hence their energy density is extremely well-measured. This means that the cosmic
neutrino background with temperature $T_\nu = (4/11)^{1/3} T_\gamma$ is well-predicted accordingly. Assuming slight heating of the neutrinos
due to $e^+e^-$ annihilation, the theoretical prediction for the effective number of neutrinos is $N_{eff} = 3.046$ \cite{Dolgov,mangano1}.
Although the neutrino energy density cannot be measured directly, it can be deduced from the CMB measurements. The knowledge of neutrino energy
density leads to the bound on $N_{eff}$.

The values of $N_{eff}$ from a combined analysis of WMAP, baryon acoustic oscillation (BAO) and Hubble constant ($H_0$) yields
$N_{eff} = 4.34^{+0.86}_{-0.88}$ (68\% CL) \cite{WMAP7B}. The further addition of the Atacama Cosmology Telescope \cite{Dunkley} and
the South Pole Telescope \cite{Keisler} lead to  $N_{eff} = 4.56 \pm 0.75$ (68\% CL) and $N_{eff} = 3.86 \pm 0.42$ (68\% CL) respectively.
Combined datasets have been used by Archidiacono \emph{et al}. \cite{Arch} to derive $N_{eff} = 4.08^{+0.71}_{-0.68}$ (95\% CL).
However, these constraints on $N_{eff}$ are derived under the assumption that the additional particles are massless. Taking into account
of the masses of two light sterile neutrinos, it has been shown that $\sum m_\nu \lesssim 0.9$ eV \cite{Hamann}, which is clearly in conflict
with the best-fit masses $m_4=$ 0.68 eV and $m_5=$ 0.94 eV \cite{Joachim} derived from neutrino experiments. Nevertheless,
this issue appears far from settled.
For instance, in a recent analysis, it has been shown that these best-fit masses are marginally
consistent with the cosmological data \cite{Giunti}. But two other recent analyses conducted by \cite{RiemerSorensen} and \cite{Xuelei}
seem to suggest $\sum m_\nu \lesssim 0.6$ eV (with $N_{eff} =3.58$) and $\sum m_\nu \lesssim 0.556$ eV (with
$N_{eff} =3.839$) respectively,
while Ref. \cite{Giusarma} obtains $\sum m_\nu \lesssim 0.32$ eV and Ref. \cite{Zhao} gives
$\sum m_\nu \lesssim 0.34$ eV.
On the other hand, the remarks made by \cite{Kev} are illuminating ---
they show that whether the cosmology of the (3+2) model is viable depends significantly on the choice of datasets included and
the ability to control the corresponding systematic uncertainties. They thus conclude that it is premature to claim that two eV sterile
neutrinos are either cosmologically favored or ruled out. As a result, it appears that we need a better understanding of the
systematic uncertainties in the analyses of cosmological datasets in order to nail down an indisputable bound on $\sum m_\nu$.

To recapitulate, we have just reviewed the current status of sterile neutrinos in the context of both particle physics and cosmology.
The forthcoming data from the PLANCK satellite will improve the bound on $N_{eff}$ to
an accuracy of\, 0.20\, at 2$\sigma$. For instance, if PLANCK sees $N_{eff} \lesssim 4.60 $, then it is unclear how to reconcile the CMB bound
 with two eV sterile neutrinos. In a worse case, if PLANCK sees $N_{eff} \lesssim 4.0$, it would appear that two eV sterile neutrinos are completely ruled out. One of the main purposes of this paper is to propose a novel mechanism, which involves MeV dark matter $p$-wave annihilation,
to accommodate two eV sterile neutrinos despite the stringent CMB bound. We will demonstrate that this mechanism is capable of accommodating two
eV sterile neutrinos even if PLANCK observes $N_{eff}$ as low as the theoretical value of 3.046.
It can also accommodate the BBN bound on the number of sterile neutrinos without introducing a large neutrino asymmetry.
We will describe this mechanism in Sec. \ref{sec:reduce} and Sec. \ref{sec:accommodate}. Then we will identify a few natural dark matter candidates that exhibit the required
properties for this mechanism in Sec. \ref{sec:relevant}.

Interestingly, a dark matter particle with precisely these required properties and lighter than 3 MeV can explain the
511 keV gamma-ray line observed by INTEGRAL \cite{Integral}. In Sec. \ref{sec:511kev}, we will show that
two eV sterile neutrinos allow this kind of dark matter to be lighter than 3 MeV, which is otherwise ruled out
by the CMB bound on $N_{eff}$ if there are no sterile neutrinos.

\section{Reducing $N_{eff}$ through MeV Dark Matter Annihilation}
\label{sec:reduce}

In this section, we present a novel mechanism to reduce $N_{eff}$ through MeV dark matter annihilation. Assuming only the three
active neutrinos, this mechanism has recently been invoked by \cite{HoScherrer} to place a lower bound on the mass of MeV-scale
dark matter. As we will see, if eV sterile neutrinos exist, the bound derived in \cite{HoScherrer} will be somewhat relaxed.

The main idea behind this mechanism is the following. An MeV dark matter particle that couples
more strongly to electrons and/or photons than to neutrinos will heat up the electron-photon plasma through pair-annihilation when it becomes
non-relativistic before its abundance freezes out. If this occurs after neutrino decoupling, then the ratio of the neutrino temperature
to the photon temperature will be reduced, leading to a decrease in $N_{eff}$. This process is similar to the heating that results from
the annihilation of electrons and positrons when they become non-relativistic.

The fact that a generic light particle, which remains in thermal equilibrium with the photons throughout
the epoch of BBN, will reduce $N_{eff}$ was first briefly discussed by \cite{KTW} in the context
of BBN. They argued that this effect may lead to an underproduction of primordial helium and deuterium.
More recently, the effect of light dark matter annihilation on BBN has been investigated by \cite{SR}.

MeV dark matter has been invoked to explain the observed 511 keV $\gamma$-rays by INTEGRAL \cite{Boehm} and
the cosmic $\gamma$-ray background at $1-20$ MeV \cite{Ahn}. It has also been found to have interesting effects on large-scale
structure \cite{Kaplinghat}. A supersymmetric model with the lightest supersymmetric particle of MeV scale has been
proposed in \cite{Hooper}. Then, in the context of supersymmetric models with gauge mediated SUSY breaking, it has been shown that MeV dark
matter serves as an example of WIMPless candidate that naturally leads to the correct relic abundance \cite{Feng}.

A constraint on MeV dark matter from core-collapsed supernovae has been derived by \cite{Fayet}. But their bound can be evaded if
the scattering cross section between the dark matter and neutrinos is negligible. Since we are interested in MeV dark matter that couples more
strongly to electrons and/or photons than to neutrinos, this bound is not relevant to our consideration.

Meanwhile, dark matter annihilation (which generates the correct relic abundance) near the epoch of recombination will
cause distortions to the CMB fluctuation spectrum and so a lower bound on the dark matter mass can be derived \cite{Padmanabhan}.
(Notice that dark matter annihilation also distorts the CMB spectrum \cite{MTW}, although the bound is much weaker.)
This effect excludes dark matter with masses $\lesssim 1-10$ GeV, which is apparently in conflict with MeV dark matter. However,
this bound is only applicable to $s$-wave annihilation, for which $\langle \sigma v \rangle$ remains constant
between the dark matter freeze-out and the epoch of recombination.  For $p$-wave annihilation which is velocity-dependent, the annihilation
rate at the recombination epoch is negligible, and so this bound can be evaded. Therefore, we require an
MeV dark matter particle with $p$-wave annihilation for our mechanism of reducing $N_{eff}$.

The following discussion is based on Ref. \cite{HoScherrer}.
Let $\chi \bar \chi$ be the pair of dark matter particle and anti-particle. (Notice that for self-conjugate and non-self-conjugate scalars,
the notation $\chi \bar \chi$ really means $\chi\chi$ and $\chi\chi^\ast$ respectively.)
Consider the case in which the dark matter annihilates entirely after the neutrinos have
decoupled in the early universe. This occurs at a temperature of $T_{d} \approx 2-3$ MeV \cite{Dolgov,Enqvist}.
The extent of the heating due to dark matter annihilation can be derived from entropy conservation \cite{Weinberg,KT}.
The total entropy before $\chi \bar \chi$ annihilation is proportional to
\begin{equation}
S = \frac{R^3}{T}\,\left(\,\rho_{e^+e^-} +  \rho_{\gamma} + \rho_{\chi \bar \chi} + p_{e^+e^-} +
p_{\gamma} + p_{\chi \bar \chi}\,\right)\,,
\end{equation}
while after $\chi \bar \chi$ annihilation, it becomes
\begin{equation}
S = \frac{R^3}{T}\,\left(\,\rho_{e^+e^-} +  \rho_{\gamma} + p_{e^+e^-} +
p_{\gamma}\,\right)\,.
\end{equation}
For a relativistic particle, the pressure is related to the density by $p = \rho/3$, so the total entropy
density can be written as \cite{KT}:
\begin{equation}
s=\frac{\rho_{\textrm{tot}}+p_{\textrm{tot}}}{T} = \frac{2\, \pi^2}{45}\, g_{*S}\, T^3\,,
\end{equation}
where $g_{*S}$ is the sum of total bosonic spin degrees of freedom and 7/8 times the total fermionic spin degrees of freedom.
Then, the total entropy equals to
\begin{equation}
S = \frac{2 \,\pi^2}{45}\, g_{*S}\, (R\,T)^3\,,
\end{equation}
which is conserved during the process where any particle species becomes non-relativistic and pair-annihilates.
Thus, the ratio of the values of $R\,T$ before and after $\chi \bar \chi$ annihilation is given by
\begin{equation}
\label{heating}
\frac{(R\,T)_i}{(R\,T)_f} = \left(\,\frac{g_{*S_f}}{g_{*S_i}}\,\right)^{1/3}\,,
\end{equation}
where $g_{*S_i}$ and $g_{*S_f}$ are respectively the values of $g_{*S}$ for the relativistic
degrees of freedom in thermal equilibrium before and after $\chi \bar \chi$ annihilation.

If the $\chi \bar \chi$ annihilation occurs
after neutrino decoupling, then the neutrinos do not share in the heating from the annihilation. This means that
$R\,T_\nu$ remains constant before and after $\chi \bar \chi$ annihilation and $T_\nu \propto R^{-1}$. But the photons and electron-positron
pairs are heated up according to Eq. (\ref{heating}).  Therefore, the ratio of $T_\nu$ to $T_\gamma$ after $\chi \bar \chi$
annihilation will be:
\bea
\label{ratf}
\left(\,\frac{T_\nu}{T_\gamma}\, \right)_{\chi \bar \chi}=
\left\{
  \begin{array}{ll}
    ~\left[\frac{(7/8)\,4 \,+ \,2}{(7/8)\,4 \,+ \,2 \,+ \,(7/8)\,g}\right]^{1/3}\,, & \hbox{for a fermionic\,$\chi$\,;} \\ \\
    ~\left[\frac{(7/8)\,4 \,+ \,2}{(7/8)\,4 \,+\, 2 \,+  \,g}\right]^{1/3}\,, & \hbox{\textrm{for a bosonic}\,$\chi$\,,}
  \end{array}
\right.
\eea
where $g$ is the total internal degrees of freedom for the $\chi \bar \chi$ pair. For a self-conjugate
scalar dark matter particle, we have $g=1$,
while for a non-self-conjugate scalar dark matter particle, we have $g=2$. Further, for a spin-1/2\,
Majorana dark matter particle which is self-conjugate,
we have $g=2$; while for a spin-1/2\, Dirac dark matter particle, we have $g=4$. Subsequent $e^+e^-$
annihilation further heats up the photons,
so the
resultant ratio of $T_\nu$ to $T_\gamma$ is given by
\begin{equation}
\label{ratfResultant}
\frac{T_\nu}{T_\gamma} = \left(\,\frac{4}{11}\right)^{1/3}\,\left(\,\frac{T_\nu}{T_\gamma}\, \right)_{\chi \bar \chi}\,.
\end{equation}

In terms of $N_{eff}$, the energy density of all relativistic neutrinos can be written as
\begin{equation}
\rho_\nu = N_{eff}\left(\frac{7}{8}\right)\,2\,\left(\frac{\pi^2}{30}\right)\,
\left(\frac{T_\nu}{T_\gamma}\right)^4 \,T_\gamma^4.
\end{equation}
The CMB observations constrain $\rho_\nu$ at fixed $T_\gamma$ and so a change in $T_\nu/T_\gamma$ with respect to its conventional value
$(4/11)^{1/3}$ (due to $e^+e^-$ annihilation) would be interpreted as a change in $N_{eff}$. But as we have shown above, $\chi \bar \chi$
annihilation reduces the value of $T_\nu/T_\gamma$ by a factor of $( T_\nu/ T_\gamma)_{\chi \bar \chi}$\,\, relative to $(4/11)^{1/3}$.
So the apparent value of $N_{eff}$ inferred from CMB observations will be a factor of $( T_\nu/ T_\gamma)_{\chi \bar \chi}^4$\, smaller than it
would without $\chi \bar \chi$ annihilation.

In this calculation, we have neglected the effect of the annihilating dark matter particle on the
expansion rate when the neutrinos drop out of thermal equilibrium (at which point the dark matter
is relativistic), as well as the possible effect of two additional sterile neutrinos.  Both
of these will increase the expansion rate and cause the neutrino decoupling temperature to increase.
We can estimate the resultant effect as follows. The Hubble rate is proportional to $\sqrt{g_{*S}}\;T^2$, while the weak interaction rate is proportional to $T^5$. The neutrino decoupling temperature $T_d$ is defined as the temperature at which the weak interaction rate drops to
roughly the same magnitude as the Hubble rate. This implies that $T_{d} \propto \left(\,g_{*S}\,\right)^{1/6}$ and so the modified neutrino
decoupling temperature will be given by
\bea
\frac{T_d^\textrm{modified}}{T_d^\textrm{original}}
= \left(\,\frac{ 2\, +\, (7/8)\, 4 \,+\,  (3+2)\,(7/8)\, 2 \,+\, (7/8)\, g }{ 2 \,+ \,(7/8) \,4 \,+ \, 3\,(7/8)\, 2 }\,\right)^{1/6}\,,
\eea
where, by writing the factor $(7/8)\, g$ in the above expression, we have assumed a fermionic dark matter. But similar arguments apply to a
bosonic dark matter. As an example, one can take $g=2$ for a Majorana dark matter. Then, we will have
$T_d^\textrm{modified}/T_d^\textrm{original} =1.07$ which indicates that the change is small.
Thus, we will ignore this effect here and in the following section.

\section{eV Sterile Neutrinos and MeV Dark Matter}
\label{sec:accommodate}

What we have considered in the last section was the effect of $\chi \bar \chi$ annihilation occurring
completely after neutrino decoupling.
In reality, it is possible that $\chi \bar \chi$ will become non-relativistic and start to annihilate
before neutrino decoupling.
In this case, the neutrinos and photons will share in the heating from
$\chi \bar \chi$ annihilation before neutrino decoupling,
while the residual $\chi \bar \chi$ annihilations after
neutrino decoupling will heat the photons alone.

Our treatment follows Ref. \cite{HoScherrer}.
After neutrino decoupling, the $\chi \bar \chi$ annihilation continues to heat up the photons relative to neutrinos until the $\chi \bar \chi$ particles drop out of thermal equilibrium. But the exercise in the last section reveals that the ratio of $T_\nu$ to $T_\gamma$ in each step of $\chi \bar \chi$ annihilation depends solely on $g_{*S_i}$ and $g_{*S_f}$. To quantify this continuous effect, we would need to consider the quantity $I(T_\gamma)$ given by
(see \cite{Weinberg} for a similar calculation):
\bea
I(T_\gamma)&\equiv& \frac{S}{(R\,T_\gamma)^3} = \frac{1}{T_\gamma^4}\,(\rho_{e^+e^-} +  \rho_{\gamma} + \rho_{\chi \bar \chi} + p_{e^+e^-} +
p_{\gamma} + p_{\chi \bar \chi}), \nonumber\\
\label{formula}
&=& \frac{11\,\pi^2}{45} + \frac {g}{2 \pi^2} \int_{x=0}^\infty \,dx\, \frac{x^2}{e^{\sqrt{x^2 + (m_\chi/T_\gamma)^2}}\pm 1} \left
(\sqrt{x^2 + (m_\chi/T_\gamma)^2} + \frac{x^2}{3\sqrt{x^2 + (m_\chi/T_\gamma)^2}}\right)\,,
\nonumber \\
\eea
where the integration variable is $x = p_\chi/T_\gamma$. The ``+" sign and ``--" sign in the integrand correspond to fermionic and
bosonic dark matter respectively.  Up to the constant factor $2 \,\pi^2/45$, the physical meaning of
$I(T_\gamma)$ is the sum of total bosonic spin degrees of freedom and 7/8 times the total fermionic spin degrees of freedom
at a given photon temperature $T_\gamma$. Notice that in the limit where
all particles are relativistic, $I$ reduces to $(2 \,\pi^2/45)\,g_{*S}$ and the integral
in Eq. (\ref{formula}) simply counts the contribution from $\chi \bar \chi$ when they become non-relativistic.

As mentioned above, the $\chi \bar \chi$ annihilation will continue to heat up photons relative to neutrinos after neutrino
decoupling until the $\chi \bar \chi$ particles drop out of equilibrium. Thus, the ratio $T_\nu$ to $T_\gamma$ due to accumulative $\chi \bar \chi$ annihilation is
\begin{equation}
\label{TfNonzero}
\left(\,\frac{T_\nu}{T_\gamma}\,\right)_{\chi \bar \chi} = \left(\,\frac{I(T_f)}{I(T_d)}\,\right)^{1/3},
\end{equation}
where $T_f$ is freeze-out temperature of the $\chi \bar \chi$ particles and is given by $T_f \sim m_\chi/20$ as a generic result \cite{KT}.
Since most of the entropy due to $\chi \bar \chi$ annihilation is deposited to the thermal background around $T \sim m_\chi/3 \gg T_f$, it
is evident from Eq. (\ref{formula}) that we can set $T_f = 0$ in Eq. \eqref{TfNonzero} as a good approximation.

\begin{figure}
\includegraphics[height=6cm, width=9cm]{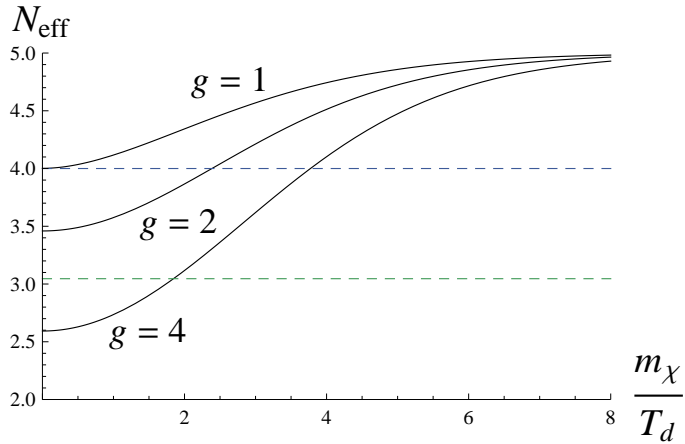}
\caption{\emph{$N_{eff}$ as a function of $m_\chi/T_d$ for two sterile neutrinos.
The curves with $g=1$, $g=2$ and $g=4$ correspond to a self-conjugate scalar, Majorana and Dirac dark matter
respectively. The two dashed lines correspond $N_{eff}=3.046$ and $N_{eff}=4.00$.}}
\label{fig:Neff}
\end{figure}

We expect each of the two eV sterile neutrinos to contribute a unity to the effective number of neutrinos.
But as a consequence of $\chi \bar \chi$ annihilation, $N_{eff}$ deduced from CMB observation would be given by
\begin{equation}
\label{Neff}
N_{eff} = 5.046\,\left(\,\frac{I(0)}{I(T_d)}\,\right)^{4/3}.
\end{equation}
The value of $N_{eff}$ against $m_\chi/T_d$
is shown in Fig. \ref{fig:Neff}, for a self-conjugate scalar ($g=1$), a spin-1/2 Majorana ($g=2$)
and a spin-1/2 Dirac ($g=4$) dark matter. The curve for a non-self-conjugate scalar ($g=2$) dark matter appears
very similar to that for a spin-1/2 Majorana dark matter and so it is not shown.

Note that the small increase in the neutrino decoupling temperature due to the $\chi \bar \chi$ particles
and the two eV sterile neutrinos, which we have chosen to neglect, will only serve to enhance
the heating of the photons relative to the neutrinos, further decreasing $N_{eff}$.

The PLANCK satellite, whose first data will be released very soon, is expected to improve the bound on $N_{eff}$ to an accuracy of 0.20 at 2$\sigma$.
If PLANCK sees $N_{eff} \gtrsim 5.0$, then our mechanism of reducing $N_{eff}$ through dark matter annihilation is not needed to accommodate
two eV sterile neutrinos. If PLANCK sees $4.0 \lesssim N_{eff} \lesssim 5.0$, one can easily read off the corresponding mass of a self-conjugate scalar, Majorana or Dirac dark matter required to accommodate two eV sterile neutrinos from Fig. \ref{fig:Neff}. If PLANCK sees $3.5 \lesssim N_{eff} \lesssim 4.0$, both Majorana and Dirac dark matter can help to accommodate two eV sterile neutrinos, while the self-conjugate scalar dark matter fails to do the job. If PLANCK sees $3.046 \lesssim N_{eff} \lesssim 3.5$, only a Dirac dark matter can do the job. In fact, a Dirac dark matter is capable of accommodating two eV sterile neutrinos even if PLANCK sees $N_{eff}$ as low as the theoretical value of 3.046.

In addition, we remark that in the unlikely event that PLANCK sees $N_{eff} \lesssim 3.046$ (in
contradiction to the standard cosmological model), then one possible resolution would be that MeV Dirac dark matter annihilation has indeed occurred to reduce $N_{eff}$. On the other hand, if PLANCK sees the highly unlikely value $N_{eff} \lesssim 2.5$, then
even a Dirac dark matter with vanishing mass cannot help to accommodate two eV sterile neutrinos. If this happens, two eV sterile neutrinos would be strongly disfavored. One may then conclude that either sterile neutrinos are not relevant for the anomalies from the short baseline neutrino experiments or some exotic physics needs to come in to rescue.

As noted above, most of the entropy due to $\chi \bar \chi$ annihilation is deposited to the thermal
background around $T \sim m_\chi/3 $. For an MeV dark matter particle, this means that most of the entropy from $\chi \bar \chi$ annihilation would have already been released before BBN starts. Thus, the value of $N_{eff}$ inferred from BBN would be roughly equal to that obtained from Eq. \eqref{Neff}. In other words, MeV dark matter $p$-wave annihilation can also help to accommodate the BBN bound on the number of sterile neutrinos. This mechanism does not require a large neutrino asymmetry, as opposed to the work by \cite{Foot,Danny}.

While we have shown that the annihilation of an MeV dark matter particle can reduce $N_{eff}$ to an acceptable
value even with two additional sterile neutrinos, we have not yet considered the observational
limits on $m_\nu$ discussed in Sec. I.  Limits on the neutrino masses are, to a first approximation, limits
on the total non-relativistic neutrino density, which is given by
\begin{equation}
\label{numass}
\rho_\nu = \sum_\nu n_\nu \,m_\nu \,,
\end{equation}
where the sum is taken over all (standard model and sterile) non-relativistic neutrinos,
and $n_\nu$ is the number density of each type of neutrino.
If all neutrinos have a common temperature, $T_\nu$, then $n_\nu$ is the same for every
neutrino and is given by
\begin{equation}
n_\nu = \left(\frac{3}{4}\right)\,2\,\left(\frac{1}{\pi^2}\right) \,\zeta(3)\, T_\nu^3\,,
\end{equation}
where $\zeta$ is the Riemann zeta function and $\zeta(3) = 1.2$.  Thus, we can rewrite Eq.
(\ref{numass}) as
\begin{equation}
\label{rhonu}
\rho_\nu = \left(\frac{3}{4}\right)\,2\,\left(\frac{1}{\pi^2}\right)\, \zeta(3)\, \left(\frac{T_\nu}{T_\gamma}\right)^3\,
T_\gamma^3 \,\sum_\nu m_\nu \,.
\end{equation}

If we have an observational bound on $\rho_\nu$ at a given $T_\gamma$, expressed as an upper bound
on $\sum {m_\nu}$, it is clear from Eq. (\ref{rhonu}) that a decrease
in $T_\nu/T_\gamma$ can compensate an increase in $\sum {m_\nu}$ and continue
to satisfy the observational bound. This implies that the apparent value of $\sum {m_\nu}$ inferred from cosmological observations
will be a factor of $( T_\nu/ T_\gamma)_{\chi \bar \chi}^3$\, smaller than it would without the MeV dark matter annihilation:
\bea
\left( \sum {m_\nu} \right)_{\textrm{cosmology}} = \left( \sum {m_\nu} \right)_{\textrm{true}}\,
\left(\frac{T_\nu}{T_\gamma}\right)^3_{\chi \bar \chi}\,.
\eea
In other words, the upper bound on $\sum m_\nu$ is increased by a factor of $(T_\gamma/T_\nu)_{\chi \bar \chi}^{3} = I(T_d)/I(0)$.
It is easy to express this factor in terms of the change in $N_{eff}$ produced by the annihilating
MeV dark matter. If $N_{eff}$ is the value deduced from CMB observations, and $N_\nu$ is
the true number of neutrinos (e.g., $N_\nu = 5.046$ in the $3+2$ model), then we see that
\bea
\left( \sum {m_\nu} \right)_{\textrm{true}}
= \left( \sum {m_\nu} \right)_{\textrm{cosmology}} \, \left(\frac{N_\nu}{N_{eff}}\right)^{3/4}\,.
\eea
So the upper bound on $\sum m_\nu$ is increased by a factor of $(N_\nu/N_{eff})^{3/4}$.
For the case of two sterile neutrinos, $N_\nu = 5.046$, and value of $N_{eff}$ (for a given $m_\chi/T_d$) can
be read off of Fig. 1.  Thus, we obtain a relationship between the dark matter annihilation
needed to satisfy upper bounds on $\sum m_\nu$, and the prediction for the observed
value of $N_{eff}$.  As an example, if the neutrino experiments give $\sum m_\nu \approx 1.6$ eV \cite{Joachim},
while cosmological observations require $\sum m_\nu \lesssim 0.9$ eV \cite{Hamann}, then
we have $(5.046/N_{eff})^{3/4} \gtrsim 1.6/0.9$, or $N_{eff} \lesssim 2.3$, which is difficult to reconcile with current
observations.  However, as noted in Sec. I, the situation with regard to upper bounds on $\sum m_\nu$
remains somewhat murky.

Our simple treatment of limits on $\sum m_\nu$ as corresponding to an upper bound
on $\rho_\nu$ alone is not entirely accurate.  The values of $m_\nu$ also
determine the temperature at which each neutrino becomes non-relativistic, and thereby
alter the process of neutrino
free-streaming, which affects large-scale structure.
In our scenario, this process is further changed because $T_\nu/T_\gamma$ is reduced
below its value in the standard model, so that a neutrino of a given mass will become non-relativistic
earlier than it would without the MeV dark matter annihilation.  However, these effects are below
the threshold for detection by current cosmological observations and would be difficult to detect
even in future surveys \cite{Jimenez}.

Clearly, our scenario differs from the standard cosmological model
in predicting a different relationship between $\rho_\nu(T_\gamma)$ when
the neutrinos are highly relativistic and $\rho_\nu(T_\gamma)$ when they are non-relativistic,
since the former scales as $(T_\nu/T_\gamma)^4$
while the latter scales as $(T_\nu/T_\gamma)^3$.  This predicted difference could be detectable
by future cosmological observations.


\section{Relevant Dark Matter Models}
\label{sec:relevant}

Since we are interested in dark matter candidates that couple more strongly to electrons and/or photons than to neutrinos, the prototypical example
could be a Dirac dark matter that interacts with the ordinary matter through an electric dipole moment (EDM) or a magnetic dipole
moment (MDM) \cite{Sigurdson}:
\bea
\mathcal{L} = \bar{\chi}\left(\,\,i\,\!\!\not\!\partial -m_{\textrm{DM}}\,\right)\chi
+ \frac{g_E \,e}{8\, m_{_\textrm{DM}}}\, \bar \chi \,\sigma^{\mu\nu}\, \tilde{F}_{\mu\nu}\, \chi
+\frac{g_M \,e}{8\, m_{_\textrm{DM}}}\, \bar \chi \,\sigma^{\mu\nu}\, F_{\mu\nu}\, \chi \,,
\eea
where $F_{\mu\nu}$ and $\tilde{F}_{\mu\nu}$ are the electromagnetic field and dual field strengths respectively.
In order to be consistent with direct detection experiments, it has been pointed out that thermal dark matter (with the correct relic abundance) acquiring EDM or MDM must have a mass less than $1-10$ GeV \cite{Banks}.

For dark matter MDM, the annihilation cross sections are \cite{Banks}:
\bea
\sigma^{\textrm{MDM}}_{\chi\bar\chi \rightarrow e^+  e^-} \,v_{\textrm{rel}} &=& \frac{(g_M\,\alpha)^2\,\pi}{4\, m^2_{_\textrm{DM}}}\,, \\
\sigma^{\textrm{MDM}}_{\chi\bar\chi \rightarrow \gamma\gamma} \,v_{\textrm{rel}} &=& \frac{(g_M^2\,\alpha)^2\,\pi}{64\, m^2_{_\textrm{DM}}}\,,
\eea
which are $s$-wave annihilations. But as mentioned earlier, $s$-wave dark matter annihilation, which leads to the correct relic abundance,
distorts the CMB fluctuation spectrum and the corresponding dark matter masses $\lesssim 1-10$ GeV are excluded. So MeV dark matter with an MDM
is not allowed.

For dark matter EDM, the annihilation cross sections are \cite{Banks}:
\bea
\sigma^{\textrm{EDM}}_{\chi\bar\chi \rightarrow e^+  e^-} \,v_{\textrm{rel}} &=& \frac{(g_E\,\alpha)^2\,\pi}{48\, m^2_{_\textrm{DM}}}\, v_{\textrm{rel}}^2 \,,\\
\sigma^{\textrm{EDM}}_{\chi\bar\chi \rightarrow \gamma\gamma} \,v_{\textrm{rel}} &=& \frac{(g_E^2\,\alpha)^2\,\pi}{64\, m^2_{_\textrm{DM}}}\,.
\eea
While $\sigma^{\textrm{EDM}}_{\chi\bar\chi \rightarrow \gamma\gamma} v_{\textrm{rel}}$ is $s$-wave, it is proportional to $g_E^4$. In the early universe, $v_{\textrm{rel}}$ is not suppressed. So it is possible that $\sigma^{\textrm{EDM}}_{\chi\bar\chi \rightarrow e^+  e^-} v_{\textrm{rel}}$, which is only proportional $g_E^2$, dominates over $\sigma^{\textrm{EDM}}_{\chi\bar\chi \rightarrow \gamma\gamma} v_{\textrm{rel}}$ and is responsible for the correct relic abundance. In this case, the bound from CMB fluctuation spectrum does not apply. So an MeV dark matter with an EDM is
a viable candidate for our mechanism of reducing $N_{eff}$.

In addition, a fermionic (Dirac or Majorana) dark matter may also acquire an anapole moment (AM) through the following
interaction \cite{HoScherrer2}:
\bea
\label{lagrangian}
\mathcal{L}_{\textrm{AM}} = \frac{g_{_A}}{\Lambda^2}\,\bar{\chi}\,\gamma^\mu\, \gamma^5\, \chi\, \partial^\nu F_{\mu\nu}\,,
\eea
where $g_{_A}$ is the coupling constant and $\Lambda$ is the cut-off scale. This interaction operator violates C and P individually, but
preserves T. Interestingly, for a Majorana fermion, both of EDM and MDM are forbidden, and the only allowed electromagnetic form factor
is the anapole moment \cite{MajoranaFermion}. At the tree-level, the annihilation cross
sections are given by \cite{HoScherrer2}:
\bea
\sigma^{\textrm{AM}}_{\chi\,\bar{\chi} \rightarrow e^+\,e^-}\,v_{\textrm{rel}} &=&
\frac{2\,g_{_A}^2\,\alpha\,m_\chi^2}{3\,\Lambda^4}\,v_{\textrm{rel}}^2 \,,\\
\label{gammagamma}
\sigma^{\textrm{AM}}_{\chi\,\bar{\chi} \rightarrow \gamma\,\gamma}\,v_{\textrm{rel}} &=& 0 \,,
\eea
which is purely $p$-wave. In contrast to EDM and MDM which interact with external electromagnetic fields, the anapole
moment interacts only with external electromagnetic currents $J_\mu = \partial^\nu F_{\mu\nu}$. But the on-shell external photons
do not lead to an electromagnetic current. This explains why we have $\sigma^{\textrm{AM}}_{\chi\,\bar{\chi} \rightarrow \gamma\,\gamma} v_{\textrm{rel}} = 0$ at the tree-level. It has been shown that dark matter with an anapole moment can generate the correct relic abundance
and be consistent with the direct detection experiments simultaneously for the mass range from MeV up to 100 GeV \cite{HoScherrer2}.
Obviously, an MeV dark matter with an anapole moment is also a viable candidate for our mechanism of reducing $N_{eff}$.

\section{511 keV Gamma-Ray Line}
\label{sec:511kev}

The INTEGRAL satellite has observed a 511 keV gamma-ray emission line from the galactic center \cite{Integral}. This emission line is expected
to arise from non-relativistic positrons annihilating with the electrons at rest in the galactic bulge. So it is important to understand the
source of these positrons.

One natural way to produce these positrons is through dark matter annihilation. For dark matter with mass larger than GeV,
the injection energies for the positrons would be too high to confine to them in the galactic bulge. Also, a substantial fraction of the
positrons might have already annihilated before reaching the required non-relativistic energies.
The idea of 1-100 MeV dark matter was then proposed by \cite{Boehm} to explain this 511 keV gamma-ray line. Later, it was shown
by \cite{Bell} that the dark matter annihilation process
$\chi \bar\chi \rightarrow e^+ e^-$ is necessarily accompanied by the electromagnetic radiative corrections
(internal bremsstrahlung) which lead to the process $\chi \bar\chi \rightarrow e^+ e^- \gamma$. What they have found was that unless
$m_\chi \lesssim 20$ MeV, the real gamma-rays emitted from the electromagnetic radiative corrections would violate the constraints on
the gamma-ray flux from COMPTEL and EGRET. By comparing the gamma-ray spectrum generated from positron inflight-annihilation and the observed
diffuse galactic gamma-ray data, the positron injection energies were further constrained to be $\lesssim$ 3 MeV \cite{Beacom}.
Recently, it has been shown that the MeV dark matter explanation is consistent with dark matter halo profiles predicted by numerical
many-body simulations for a Milky Way-like galaxy \cite{Cline}.

For an MeV dark matter, the only kinematically allowed annihilation modes are $e^+ e^-$, photons and neutrinos. But to avoid the direct gamma-ray
constraint such as the cosmic gamma-ray background \cite{Zhang} and to maintain the required positron production rate, the annihilations into photons and neutrinos are postulated to be suppressed. Therefore, the current consensus is that in order to simultaneously explain this 511 keV gamma-ray line and produce the correct relic abundance, we need a dark matter with mass $m_\chi \lesssim 3$ MeV which annihilates primarily into $e^+ e^-$ through $p$-wave \cite{Boehm}.\footnote{See, however, \cite{Lingenfelter} for possible astrophysical arguments against the dark matter interpretation
of the 511 keV gamma-ray line.}

The MeV dark matter required could be a scalar or fermion which annihilates primarily into $e^+ e^-$ through a new
light gauge boson \cite{Boehm,BoehmFayet}. It could also be a dark matter with an EDM or anapole moment as mentioned in Sec. \ref{sec:relevant}.
In other words, the dark matter, which can help to accommodate two eV sterile neutrinos and is lighter than 3 MeV, is precisely the one required to
explain the 511 keV gamma-ray line observed by INTEGRAL.


One the other hand, as shown in \cite{HoScherrer}, for any dark matter that couples more strongly to electrons and/or photons than to neutrinos and annihilates through $p$-wave, the lower mass bound set by $N_{eff}$ inferred from CMB is at least $m_\chi > 3$ MeV. This mass bound applies to any dark matter that may be a self-conjugate scalar, non-self-conjugate scalar, Majorana fermion or Dirac fermion. So the possibility of explaining the
511 keV gamma-ray line by dark matter with $m_\chi \lesssim 3$ MeV is ruled out by the lower mass bound derived in \cite{HoScherrer}.
However, the derivation in \cite{HoScherrer} assumed no sterile neutrinos. With two eV sterile neutrinos, it is evident from
Fig. \ref{fig:Neff} that the lower mass bound can easily be relaxed. This reopens the window for $m_\chi \lesssim 3$ MeV.

From Fig. \ref{fig:Neff}, if we choose a reasonable lower bound on $N_{eff}$ to be $N_{eff} \gtrsim 3.046$, then there is no lower mass bound for a self-conjugate/non-self-conjugate scalar or Majorana dark matter, while a Dirac dark matter must have $m_\chi \gtrsim 4$ MeV. So a Dirac dark matter, including the one with EDM or anapole moment, cannot be a candidate to explain the 511 keV gamma-ray line. But a self-conjugate scalar or Majorana dark matter may still be viable depending on the value of $N_{eff}$ to be measured by PLANCK.
For instance, if PLANCK measures $4.0 \lesssim N_{eff} \lesssim 4.2$, only the self-conjugate scalar is a viable candidate. If
PLANCK measures $3.5 \lesssim N_{eff} \lesssim 3.7$, only the Majorana dark matter is a viable candidate. Otherwise, it becomes impossible to
explain the 511 keV gamma-ray line by dark matter annihilation.

Finally, we would like to compare the models that consist of a Majorana dark matter particle for explaining the 511 keV gamma-ray line. So far, all the proposed models require the existence of a new light gauge boson and need to postulate that the dark matter annihilations into photons and neutrinos are somehow suppressed \cite{Boehm,Hooper}. This ensures that the Majorana dark matter annihilates primarily into $e^+ e^-$. On the other hand, a Majorana dark matter with an anapole moment exhibits all of these required properties in a simpler and more natural manner \cite{HoScherrer2}.
Firstly, Eq. \eqref{lagrangian} dictates that the dark matter anapole moment couples to an external electromagnetic current. The Majorana dark matter annihilates through the Standard Model photons. So no new gauge boson is required. Secondly, since the photons do not couple to the neutrinos, the dark matter annihilation into neutrinos is naturally suppressed. Thirdly, according to Eq. \eqref{gammagamma}, we have \,$\sigma^{\textrm{AM}}_{\chi\,\bar{\chi} \rightarrow \gamma\,\gamma}\,v_{\textrm{rel}} = 0$\, at the tree-level, and so the dark matter annihilation into photons is naturally suppressed as well. The dark matter annihilation into $e^+ e^-$ is the predominant mode and is $p$-wave.
Therefore, an MeV Majorana dark matter with an anapole moment is the simplest and most natural candidate to explain the 511 keV gamma-ray line.

\section{Conclusions}

In this article, we presented a mechanism for reducing $N_{eff}$ through MeV dark matter $p$-wave annihilation. This occurs if
the MeV dark matter couples more strongly to electrons and/or photons than to neutrinos. Consequently, we showed that this mechanism
can help to accommodate two eV sterile neutrinos even if PLANCK observes $N_{eff}$ as low as the theoretical value of 3.046.
It can also help to accommodate the BBN bound on the number of sterile neutrinos without introducing a
large neutrino asymmetry. This mechanism also weakens the upper bound on the neutrino masses.  Further,
our model (or indeed any model which changes the ratio between $T_\nu$ and $T_\gamma$) produces a different relationship between
$\rho_\nu(T_\gamma)$ when the neutrinos are highly relativistic and $\rho_\nu(T_\gamma)$ when they are non-relativistic, since the former
scales as $(T_\nu/T_\gamma)^4$ while the latter scales as $(T_\nu/T_\gamma)^3$.  This modification to the standard cosmological model
could become apparent (or ruled out) as the observational data improve.

We pointed out that a natural dark matter candidate which exhibits the desirable properties for this mechanism could be the
one with an EDM or anapole moment. Interestingly, in order to explain the 511 keV gamma-ray line observed by INTEGRAL, we need
precisely a dark matter that enables this mechanism and is lighter than 3 MeV. We showed that while this kind of dark matter lighter than
3 MeV is ruled out by the CMB bound on $N_{eff}$ if only active neutrinos are considered, two eV sterile neutrinos reopen the
window for $m_\chi \lesssim 3$ MeV. Finally, we argued that an MeV Majorana dark matter with an anapole moment is the simplest and most natural
candidate to explain the 511 keV gamma-ray line, and at the same time, can help to accommodate two eV sterile neutrinos through its
purely $p$-wave annihilation into $e^+ e^-$.

It is remarkable that the solution for the anomalies in LSND and MiniBooNE (two eV sterile neutrinos) is somehow related to
the solution for the 511 keV gamma-ray observed by INTEGRAL (MeV dark matter).  On the one hand, we need MeV dark matter $p$-wave annihilation
to reduce $N_{eff}$ so as to accommodate two eV sterile neutrinos despite the BBN and CMB bounds. On the other hand, two eV sterile
neutrinos are useful in allowing a dark matter of exactly the same kind to be lighter than 3 MeV. We thus conclude that eV sterile neutrinos
and MeV dark matter are complementary to each other.

\acknowledgments
C.M.H. and R.J.S. were supported in part by the Department of Energy (DE-FG05-85ER40226).

{}

\end{document}